\documentclass[11pt]{article}
\usepackage{amssymb,cite,graphicx}
\newcommand\blank[1]{}

\newcommand\toline[1]{--#1}

\newcommand{\fract}[2]{{\textstyle\frac{#1}{#2}}}

\newcommand{\CS}{{\cal S}}

\newcommand\ZZ{{\mathbb Z}}

\newcommand{\balpha}{\alpha\kern -6.7pt\alpha}
\newcommand{\bbalpha}{\alpha\kern -4.95pt\alpha}
\newcommand\phup{^{\phantom p}}

\newcommand{\CaC}{{\cal C}}
\newcommand{\CH}{{\cal H}}

\newcommand{\CN}{{\cal N}}

\newcommand{\CQ}{{\cal Q}}

\newcommand\eq{\begin{equation}}
\newcommand\en{\end{equation}}
\newcommand\bea{\begin{eqnarray}}
\newcommand\eea{\end{eqnarray}}

\newcommand{\One}{{\hbox{{\rm 1{\hbox to 1.5pt{\hss\rm1}}}}}}
\renewcommand{\One}{{\mathbb 1}}
\renewcommand{\One}{{\rm 1\!\!1}}

\begin{document}
\begin{titlepage}
\vskip 0.5cm
\begin{flushright}
DCPT/01/37  \\
{\tt hep-th/0104119}\\
\end{flushright}
\vskip 1.8cm
\begin{center}
{\Large \bf Supersymmetry and the spontaneous\\[8pt] breakdown of ${\cal PT}$
symmetry}
\end{center}
\vskip 1.2cm
\centerline{Patrick Dorey%
\footnote{\tt p.e.dorey@durham.ac.uk},
Clare Dunning%
\footnote{\tt tcd1@york.ac.uk}
and Roberto Tateo%
\footnote{\tt roberto.tateo@durham.ac.uk}
}
\vskip 0.9cm
\centerline{${}^{1,3}$\sl\small Dept.~of Mathematical Sciences,
University of Durham, Durham DH1 3LE, UK\,}
\vskip 0.2cm
\centerline{${}^2$\sl\small 
Dept.~of Mathematics, University of York, York YO10 5DD, UK }
\vskip 1.25cm
\vskip 0.9cm
\begin{abstract}
\vskip0.15cm
\noindent
The appearances of complex eigenvalues in the
spectra of  ${\cal PT}$-symmetric quantum-mechanical systems are
usually associated with a spontaneous breaking of ${\cal PT}$.
In this letter we discuss a family of models for which
this phenomenon is also linked with an explicit breaking of supersymmetry. 
Exact level-crossings are located, and connections 
with $\CN$-fold supersymmetry and
quasi-exact solvability in certain special cases are pointed out.
\end{abstract}
\end{titlepage}
\setcounter{footnote}{0}
\def\thefootnote{\fnsymbol{footnote}}
\noindent{\bf 1.}
In a recent paper \cite{DDT3}, we discussed the spectra of the
following family of 
${\cal PT}$-symmetric eigenvalue problems:
\eq
\Bigl[-\frac{d^2}{dx^2}-(i x)^{2M}
-\alpha (i x)^{M-1}+ \frac{l(l{+}1)}{x^2}
\Bigr]\psi(x)=\lambda\,\psi(x)\, ,
{}~~~\psi(x) \in L^2(\CaC)\,.
\label{PTg}
\en
The wavefunction $\psi(x)$ is required to be square-integrable on the
contour $\CaC$. For $M<2$ this can be the 
real axis, while for $M\ge 2$, it should be distorted down into the complex
$x$-plane, as in \cite{BB} for the special case $\alpha=l(l{+}1)=0$.

The principal interest in ${\cal PT}$-symmetric
problems lies in the fact that, despite not
being in any obvious sense Hermitian, they often appear to have
spectra which are entirely real 
\cite{BZJ,BB,BBM,DT,Mez,ZCBR,BDMS,BBMSS,BW,Mez1,Zn}.
In appendix~B of \cite{DDT3}, a {\em proof} of this property was given for
the class of Hamiltonians (\ref{PTg}), drawing on ideas from the
so-called `ODE/IM correspondence'
\cite{DTa,BLZa,DTb,Sc,DDTr} (see also
\cite{voros}). More precisely, it was shown that, for $M$, $\alpha$
and $l$ real and $M>1$, the spectrum of (\ref{PTg}) is
\bea
\bullet&\! \mbox{~~{\em real}~~~~ if}&\alpha<M+1+|2l{+}1|~; 
\label{rres}\\[3pt]
\bullet&\!\! \mbox{{\em positive}~ if}&\alpha<M+1-|2l{+}1|~. 
\label{pres}
\eea
Referring to figure~\ref{regions}, reality was
proved for $(\alpha,l)\in B\cup C\cup D$, and positivity for 
$(\alpha,l)\in D$.
\[
\begin{array}{c}
\refstepcounter{figure}
\label{regions}
\includegraphics[width=0.3\linewidth]{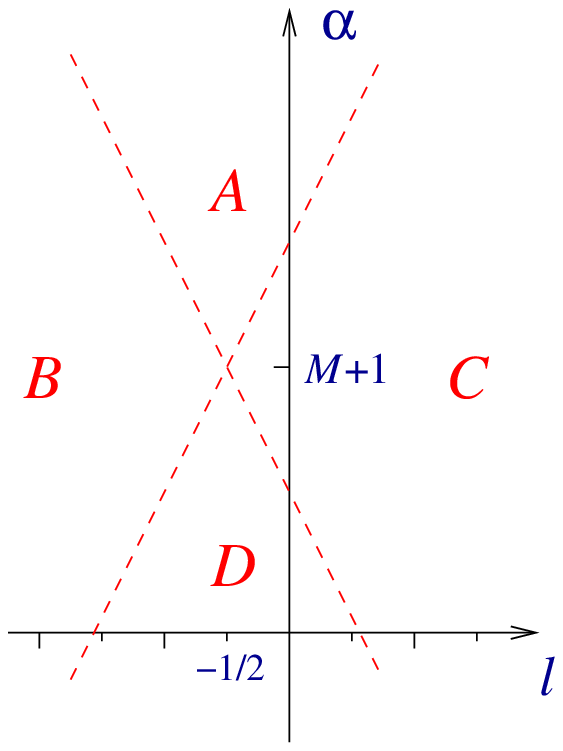}
\\[-5pt]
\parbox{0.52\linewidth}{
\raggedright
{\small Figure \ref{regions}: The initial `phase diagram' at fixed $M$.
}}
\end{array}
\]
Special cases of this result were the subject of
a number of previous conjectures: for
$\alpha{=}l{=}0$, $M{=}3/2$ in 
\cite{BZJ}; for $\alpha{=}l{=}0$ in \cite{BB}; and for $\alpha{=}0$, $l$
small in \cite{DTb}. 
More unexpected was the possible loss of reality in
the region $A$. The condition (\ref{rres}) arose for technical reasons in
the proof in \cite{DDT3}, though it was verified that at some (but not
all) points inside $A$, reality did indeed break down. But the reason for
this breakdown, and the physical significance -- if any -- of the
boundaries of the region $A$, were left obscure.

In this note we show that the appearances of complex eigenvalues as 
the lines $\alpha=M+1+|2l{+}1|$ are crossed are directly related to the
fact that on these lines the problem (\ref{PTg}) can be reformulated
in terms of a ${\cal PT}$-symmetric version of supersymmetric quantum mechanics,
and exhibits level-crossing. Elsewhere in the phase
diagram, supersymmetry is explicitly broken and
the level-crossing is lifted. Moving into the region $A$, this lifting
occurs through the creation of a pair of complex-conjugate energy levels. 
Since the presence of complex energy levels indicates
the spontaneous breaking of ${\cal PT}$ symmetry~\cite{BB}, 
we see that the model shows an interesting interplay between the
two different ways in which  a symmetry can be broken.

\medskip\noindent{\bf 2.}
To streamline some
formulae, we replace $x$ by $x/i$ in (\ref{PTg}) by setting
$\Phi(x)=\psi(x/i)$, so that the eigenproblem becomes
\eq
\Bigl[-\frac{d^2}{dx^2}+ x^{2M}
+\alpha  x^{M-1}+ \frac{l(l{+}1)}{x^2}
\Bigr]\Phi(x)=-\lambda\,\Phi(x)\, ,
{}~~~\Phi(x) \in L^2(i\,\CaC)\,.
\label{PT}
\en
We now specify the quantization contour more precisely: $i\,\CaC$ starts
and ends at $|x|=\infty$, joining the (Stokes) sectors $\CS_{-1}$ 
and $\CS_1$, where
\eq
\CS_k\,=\,\left\{\,x\,:\, \left|\,\arg(x)-
\frac{2\pi k}{2M{+}2}\,\right|<\frac{\pi}{2M{+}2}\,\right\}~.
\en
This is illustrated in figure \ref{contour}.
\[
\begin{array}{c}
\refstepcounter{figure}
\label{contour}
\includegraphics[width=0.42\linewidth]{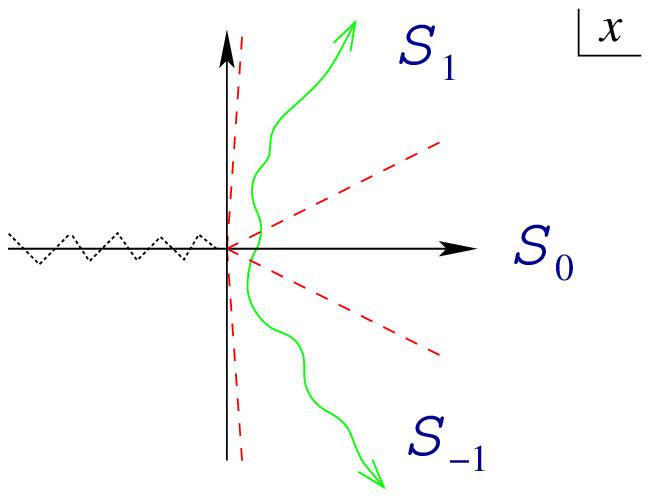}
\\[-5pt]
\parbox{0.65\linewidth}{
\raggedright
{\small Figure \ref{contour}: The quantization contour $i\,\CaC$
(arrowed line).
}}
\end{array}
\]
It will also be convenient to adopt a new set of coordinates on the
$(\alpha,l)$ plane, by setting
\eq
\alpha_{\pm}=\frac{1}{2M{+}2}\left[\,\alpha-M-1\pm(2l{+}1)\,\right]
\en
so that
\eq
\alpha=(M{+}1)(1+\alpha_++\alpha_-)~~,~~~
2l+1 = (M{+}1)(\alpha_+-\alpha_-)~.
\en
The domain boundaries -- the dotted lines on figure~\ref{regions} -- are
the lines $\alpha_{\pm}=0$.
On the line $\alpha_+=0$, $\alpha=M{-}2l$ and the problem (\ref{PT}) 
factorises as
\eq
\CQ_+\CQ_-
\Phi(x)=-\lambda\,\Phi(x)~,
\quad\mbox{with}~~\CQ_{\pm}=
\Bigl[\pm\frac{d}{dx}+ x^{M}-\frac{l}{x}\Bigr]~.
\label{spro}
\en
Such a factorisation is usually taken to signal a relationship with 
supersymmetry~\cite{Wit,Shif}. Indeed,  it is immediately seen that
(\ref{spro}) has a $\lambda=0$ eigenfunction in $L^2(i\CaC)$\,: 
\eq
\Psi(x)=x^{-l}\exp\left(\fract{1}{M{+}1}x^{M+1}\right)~,\quad \CQ_-\Psi(x)=0\,,
\en
which can be
interpreted as having unbroken supersymmetry.
All other eigenfunctions are paired with those of the SUSY partner
Hamiltonian $\widehat\CH=\CQ_-\CQ_+$\,.
This is found by
replacing
$(\alpha,l)=(M{-}2l,l)$ by $(\widehat\alpha,\widehat l)=(-M{-}2l,l{-}1)$,
or $(\alpha_+,\alpha_-)
=(0,-\frac{2l{+}1}{M{+}1})$ 
by
$(\widehat\alpha_+,\widehat\alpha_-)=(-1,-1{-}\frac{2l{-}1}{M{+}1})
=(-1,\alpha_-{-}\frac{M{-}1}{M{+}1})$.
(A rather different point of view on supersymmetry in ${\cal PT}$-symmetric
quantum mechanics was taken in \cite{ZCBR}, where the action of the supersymmetry 
generators $\CQ_{\pm}$
was supplemented by ${\cal T}$, resulting in a set-up with two states of zero
energy, rather than one.)
Replacing $-l$ by $l{+}1$ gives a $\lambda=0$ eigenfunction on the line
$\alpha_-=0$.
Thus the boundaries between the regions on figure \ref{regions} are
picked out by the presence of a (supersymmetric) zero-energy state in the
spectrum of the model. 
One would normally expect this to be the ground state, 
and indeed this is
the case on the boundary of $D$. However, level-crossing 
means that $\Psi(x)$ is only the ground state on the boundary of 
$A$ for $\alpha<M{+}3$. There is no contradiction with the usual theorems
of supersymmetric quantum mechanics, since the problem under discussion is
not Hermitian.

To verify that level-crossing does occur, we consider
$T(0)$, where $T(-\lambda)$ is a spectral determinant which
vanishes if and only if 
(\ref{PT}) has a solution, square-integrable
on $i\,\CaC$, at that value of $\lambda$.
In \cite{DDT3}, an expression for $T(0)$ was found. In the `light-cone'
coordinates $\alpha_{\pm}\,$, this is
\eq
T(-\lambda,\alpha_+,\alpha_-)|\phup_{\lambda=0}=
\Bigl(\fract{M{+}1}{2}\Bigr)^{1+\alpha_++\alpha_-}
\frac{2\pi}
{\Gamma\left(-\alpha_+\right)
\Gamma\left(-\alpha_- \right)}~.
\label{Tzero}
\en
As expected from supersymmetry, $T|_{\lambda=0}$ is identically zero when
either
$\alpha_+$ or $\alpha_-$ vanishes, 
on account of the presence of the state $\Psi$ in the
spectrum.
Level-crossing will occur when a further level passes through zero, but the
presence of $\Psi$ makes this hard to detect from an examination of
$T(-\lambda,\alpha_+,\alpha_-)|_{\lambda=0}$ alone. 
It is tempting suppose that the level-crossings happen
at the double zeroes of $T|_{\lambda=0}$. Tempting, but wrong, for reasons which may
become clearer when figures \ref{trans} and \ref{ntrans}
 below are examined. Better is to consider the
SUSY partner potential to (\ref{spro}), which is
isospectral to it save for the elimination of
the state $\Psi$. 
Substituting the values of $\widehat\alpha_+$ and $\widehat\alpha_-$ into
(\ref{Tzero}),
\eq
T(-\lambda,\widehat\alpha_+{=}-1,\widehat\alpha_-)|\phup_{\lambda=0}=
\Bigl(\fract{M{+}1}{2}\Bigr)^{\widehat\alpha_-}
\frac{2\pi}{\Gamma\left(-\widehat\alpha_-\right)}=
\Bigl(\fract{M{+}1}{2}\Bigr)^{\widehat\alpha_-}
\frac{2\pi}{\Gamma\left(\frac{M{-}1}{M{+}1}-\alpha_-\right)}~.
\label{Tdzero}
\en
Level-crossings with the state $\Psi$
are indicated by 
simple zeroes of (\ref{Tdzero}), and are at
\eq
(\alpha_+,\alpha_-)=(0,n+\fract{M{-}1}{M{+}1})~,
\qquad
n=0,1,\dots\,.
\label{levcr}
\en
 (Had we looked instead for double zeroes of
(\ref{Tzero}), we would have predicted --~incorrectly~-- the values
$\alpha_-=n$\,.)
Swapping $\alpha_+$ and $\alpha_-$ throughout gives the level-crossings on the
line $\alpha_-=0$.
Note that these level-crossings are {\em exact} -- the state $\Psi$ is
protected by supersymmetry, and cannot mix with any other state, even as the
level-crossing value of $l$ is approached. 
However, as soon as the supersymmetric line is left, the protection is lost and
mixing does occur. This is the key point, showing why the boundaries of the
region $A$ are of more than just technical significance. Since the model is ${\cal
PT}$-symmetric, all energies are either real or occur in complex-conjugate
pairs~\cite{BB}, and a real energy can only become complex if it first pairs off
with another real energy.  What we have just shown is that the hidden
supersymmetry of
the theory on the lines $\alpha_+=0$ and $\alpha_-=0$ affords
a mechanism for this pairing-off to occur, by permitting 
eigenvalues to become exactly-degenerate.

\medskip\noindent{\bf 3.}
To verify this picture, we report some numerical data for 
$M=3$. This value is chosen principally for convenience, since at $M=3$ the
fifth spectral equivalence discussed in \cite{DDT3} allows the `lateral'
eigenvalue problem (\ref{PT}), defined on the contour $i\,\CaC$, to be mapped
onto a radial problem defined on the positive real axis. 
(Note that this mapping does not
rule out the appearance of complex eigenvalues of (\ref{PT}), since at 
such points the corresponding radial problem has `irregular' boundary
conditions at the origin and is not self-adjoint.)
The merit of the radial problem is that it is straightforward to treat
numerically, using (for example) the {\sc Maple} code in appendix~A of
\cite{DDT3}. 
\[
\begin{array}{ccc}
\refstepcounter{figure}
\label{trans}
\!\!\!\!\!\!
\includegraphics[width=0.32\linewidth]{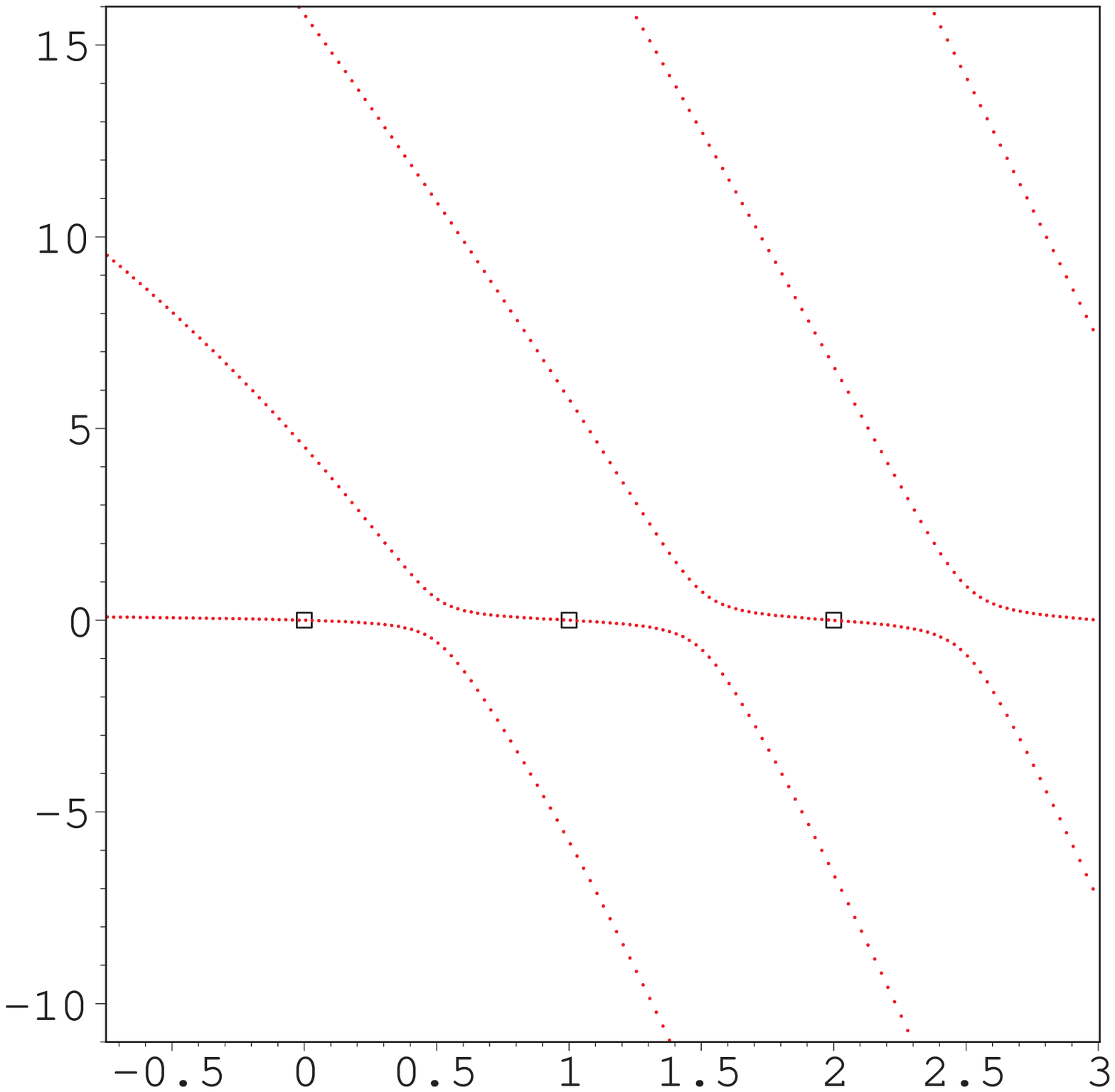}\,&
\includegraphics[width=0.32\linewidth]{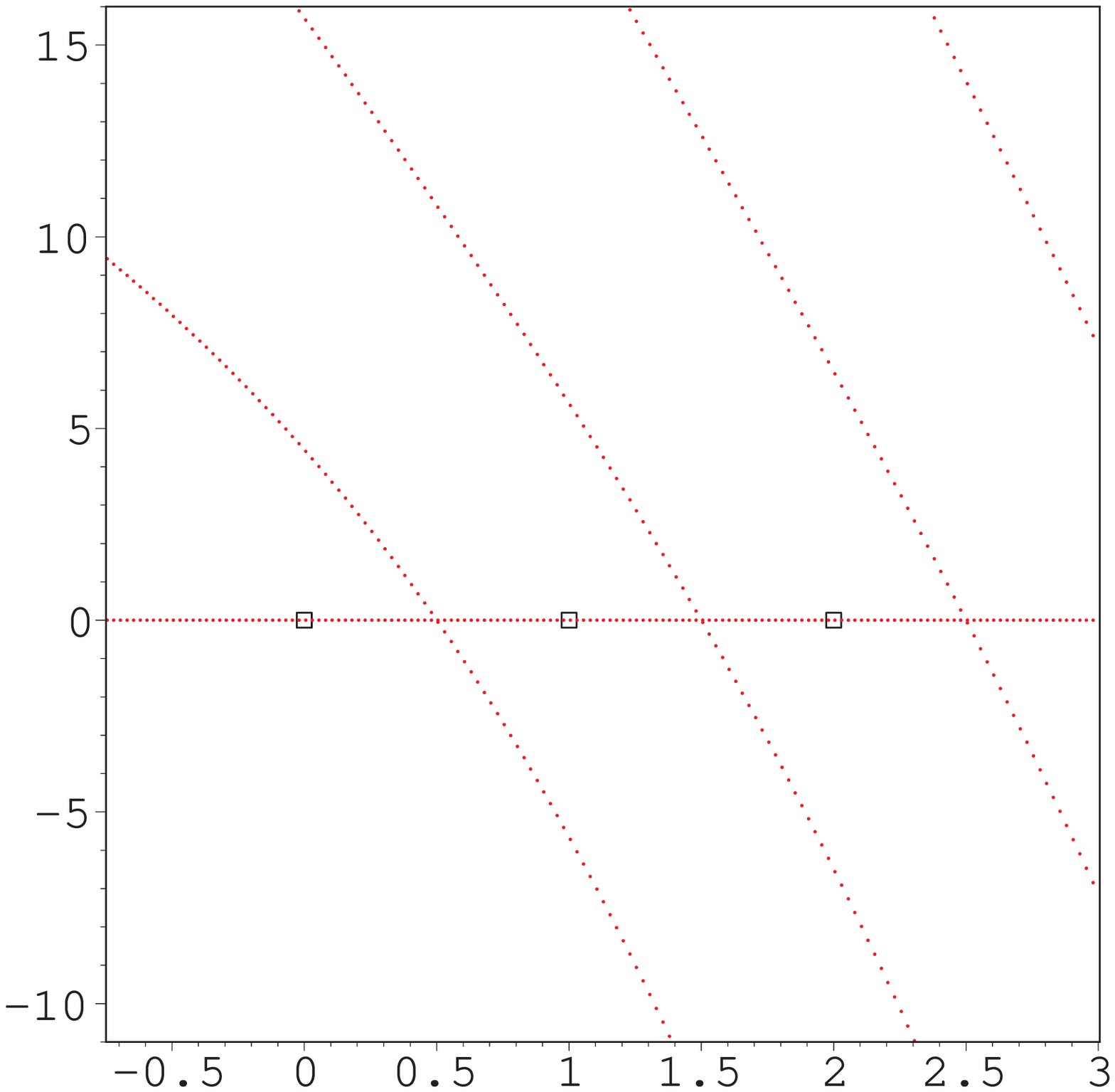}\,&
\includegraphics[width=0.32\linewidth]{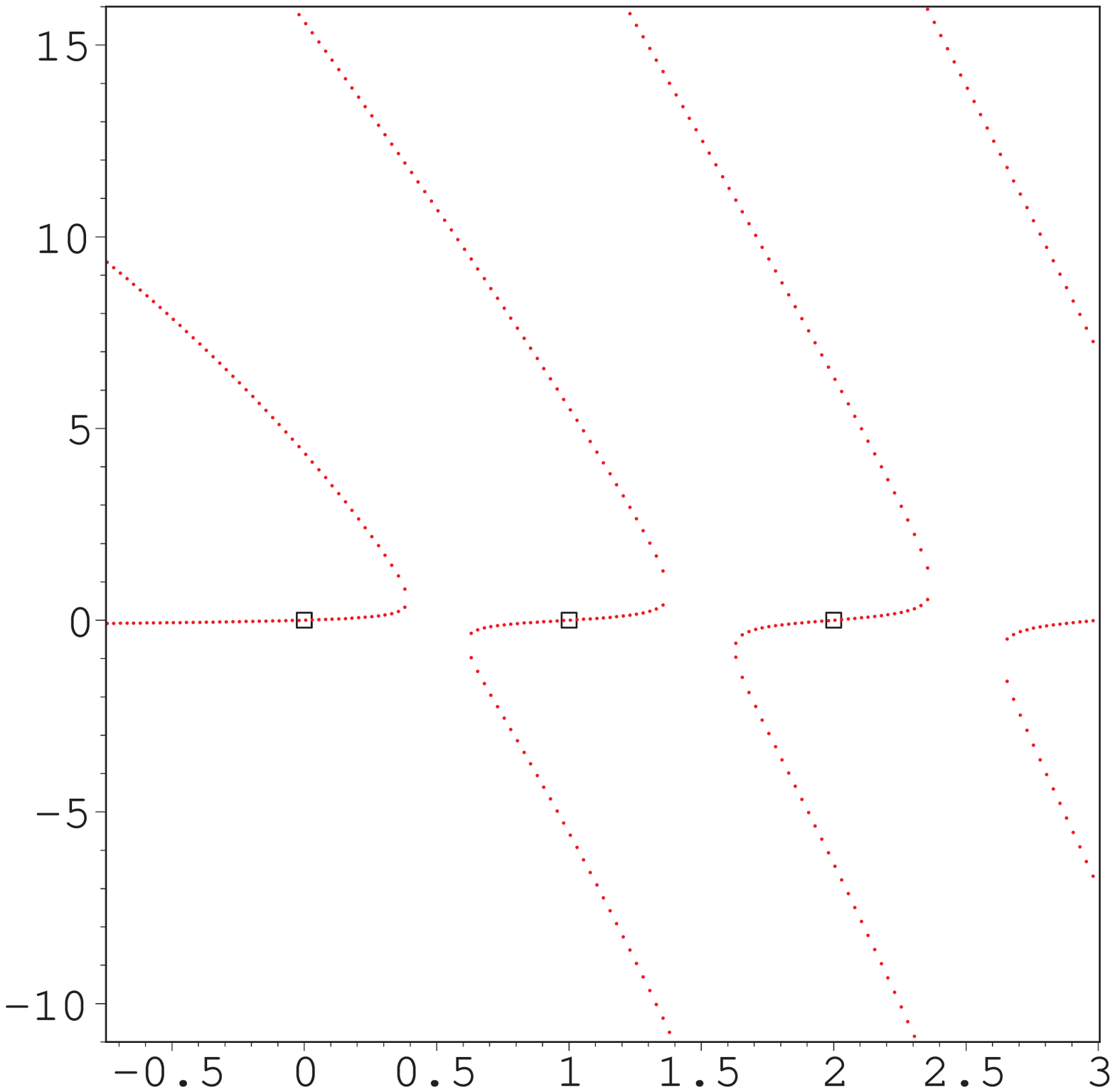}
\\[3pt]
\!\!\!\!\!\!
\parbox{0.32\linewidth}{\small~~~~~\,$\alpha_+=-0.01$}&
\parbox{0.32\linewidth}{\small~~~~~\,$\alpha_+=0$}&
\parbox{0.32\linewidth}{\small~~~~~\,$\alpha_+=0.01$}
\end{array}
\]
\[
\parbox{0.8\linewidth}{\small\raggedright
Figure~\ref{trans}: The appearance of complex eigenvalues 
as the supersymmetric line $\alpha_+=0$ is crossed, for $M=3$. 
The eigenvalues are
plotted against $\alpha_-$\,,
with the points
$\alpha_-\in\ZZ^+$, $E=0$
indicated by boxes; for $\alpha_+=0$, these are double zeroes of
$T(-\lambda,\alpha_+,\alpha_-)$\,.}
\]

\medskip

In figure~\ref{trans} we show the real energy levels along the supersymmetric
line $\alpha_+=0$, and along two lines just either side of it. On the third of
these plots -- the only one partially inside the region $A$ -- pairs of
eigenvalues join and become complex in the
neighbourhoods of the level-crossing points. By contrast, on the first plot,
which lies entirely inside regions covered by the proof of
\cite{DDT3}, the spectrum remains entirely real and the effect of
supersymmetry-breaking is seen in the replacement of
level-crossing by level-repulsion.

It might seem surprising that we are observing level-crossing in such a simple
quantum-mechanical system. Again, the non-Hermiticity of the problem provides
the explanation. As the level-crossing point is approached, not only the two
eigenvalues but also their eigenfunctions become equal. (This is clear since
eigenfunctions of (\ref{PT})
are uniquely characterised by the values of $\alpha_{\pm}$
and $E$, and their decay in the sector $\CS_{-1}$.)
The Hamiltonian thus ceases to be completely diagonalisable, a situation which
can be understood by considering the following $2\times 2$ matrix, with
eigenvalues $\pm\sqrt{\delta}$\,:
\eq
\CH\phup_{\delta} = 
\left(
\begin{array}{cc}
0&1{+}\delta\\
\delta&0
\end{array}
\right)~;\qquad
\CH\phup_{\delta}
\left( \begin{array}{c} 1\\ \pm\sqrt{\delta} \end{array} \right)
= \pm\sqrt{\delta}
\left( \begin{array}{c} 1\\ \pm\sqrt{\delta} \end{array} \right)~.
\en
The value of $\delta$ corresponds to the distance from the
level-crossing in the supersymmetry-breaking direction. 
(The addition of $\mbox{diag}(-\eta,0)$ to $\CH\phup_{\delta=0}$
would instead model a perturbation which
preserves supersymmetry, leaving the theory on the line $\alpha_+=0$.)
At $\delta=\eta=0$ the two eigenvectors coincide. For this reason, the
reference at the end of the last section to
a {\em degeneracy} in the model at the level-crossing should be treated with
caution -- better maybe to say that the eigenproblem itself
has become singular. The two
eigenvalues become complex for $\delta$
negative, so the transition to complex eigenvalues can be traced to
the singular nature of the eigenproblem at the level-crossing.

(To avoid confusion, we should stress that we are talking only of the `bosonic'
part of a fully supersymmetric problem here -- the full
SUSY QM system involves both $\CH(\alpha_+,\alpha_-)$ and 
$\CH(\widehat\alpha_+,\widehat\alpha_-)$, and the
singularity of the eigenproblem at the level-crossing
is reflected in the fact that the $\lambda=0$
degeneracy jumps from $1$ to $2$ there, rather than from $1$ to $3$
as would naively have been expected.)

The lines $\alpha_{\pm}=0$
are not the only ones along which there is a `protected' zero-energy level in
the spectrum of the system -- from the formula (\ref{Tzero}), 
the same is true  of all the lines $\alpha_{\pm}=n$,
$n\in\ZZ^+$. At least for $M=3$, we can understand this as being due to a
hidden $\CN$-fold supersymmetry in the model \cite{DDT3,nfold}.
In figure
\ref{ntrans} we illustrate how further complex levels are created as these 
lines are crossed. 

\vskip 2pt
\[
\begin{array}{ccc}
\refstepcounter{figure}
\label{ntrans}
\!\!\!\!\!\!
\includegraphics[width=0.32\linewidth]{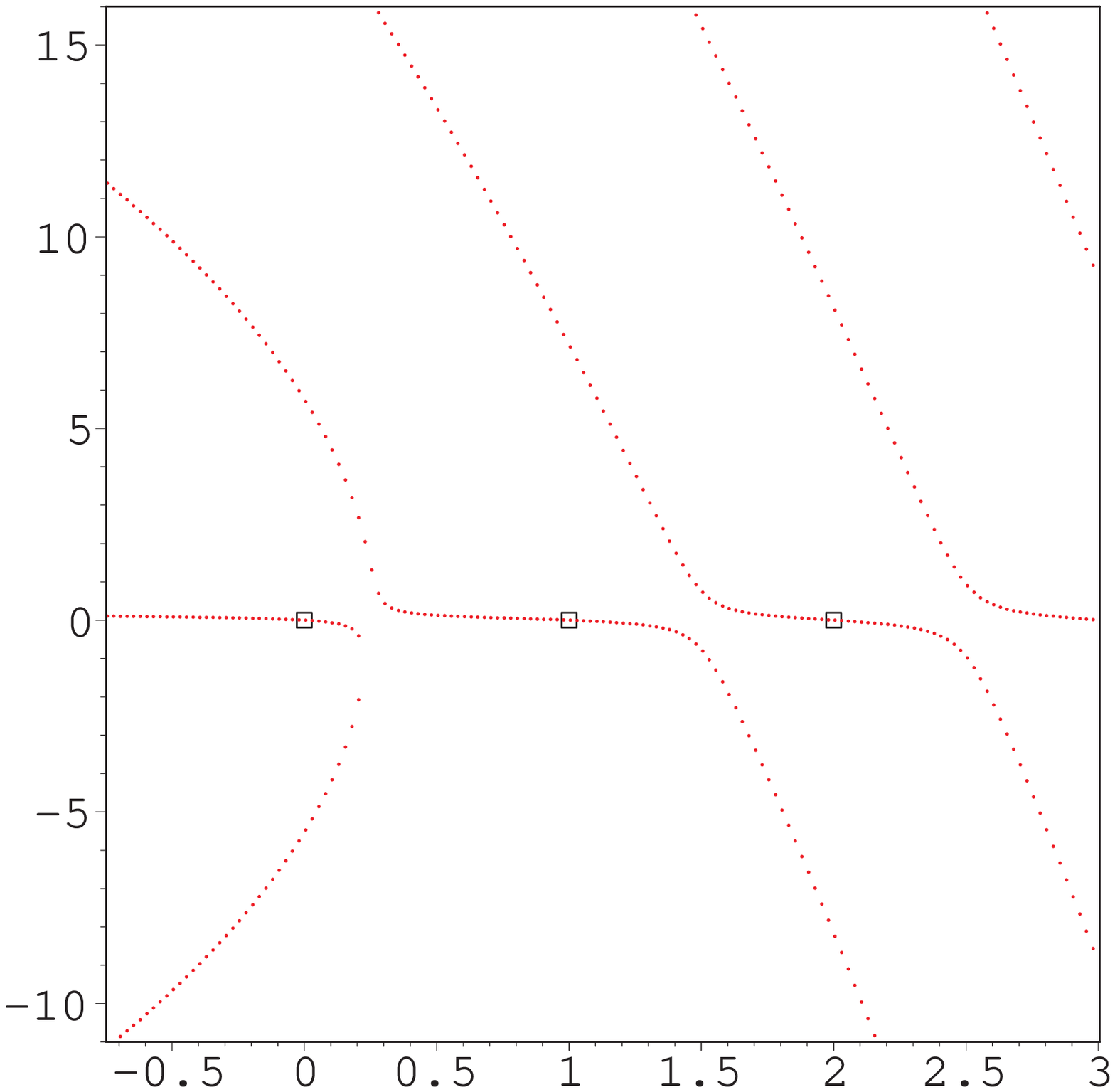}\,&
\includegraphics[width=0.32\linewidth]{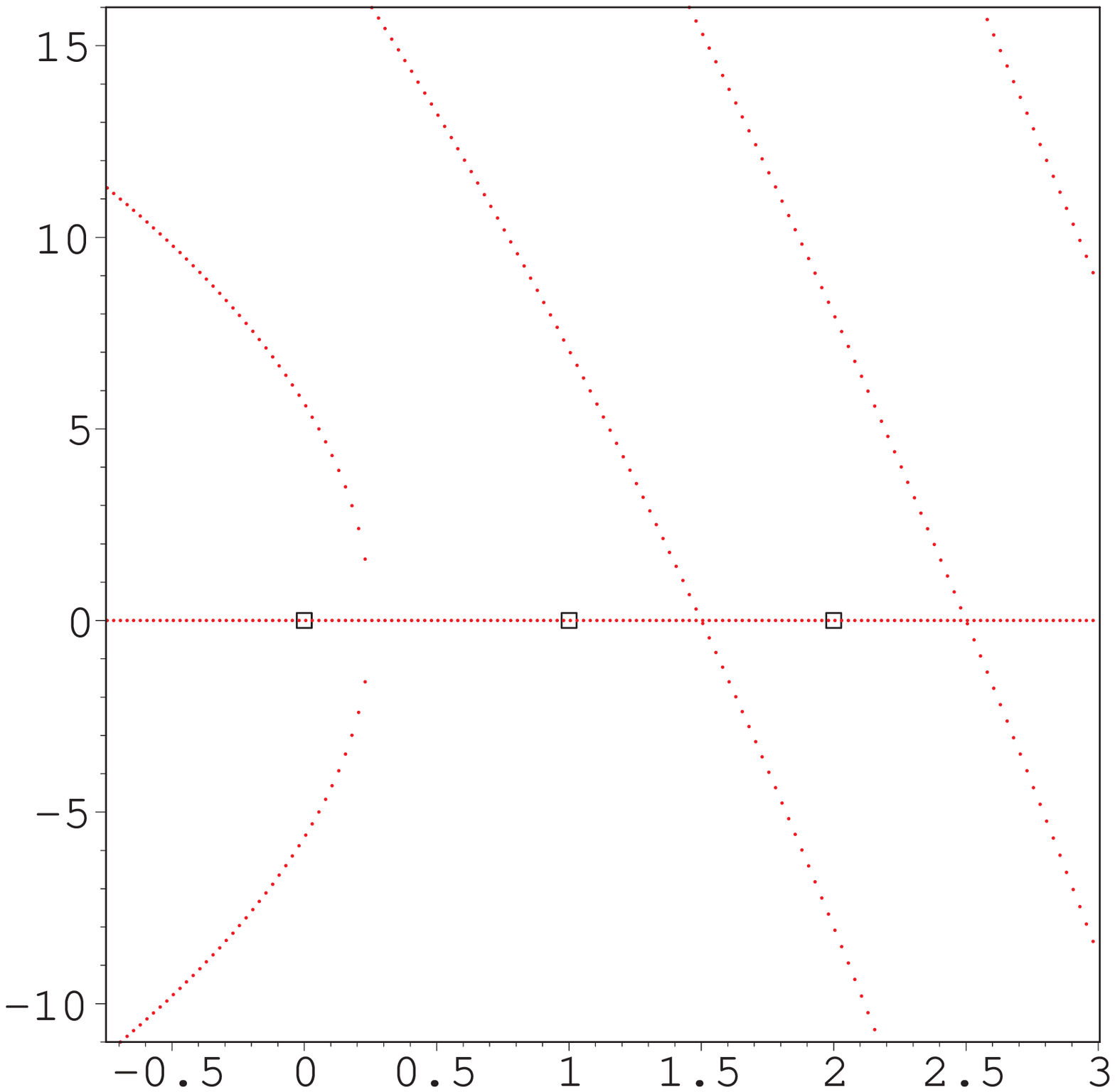}\,&
\includegraphics[width=0.32\linewidth]{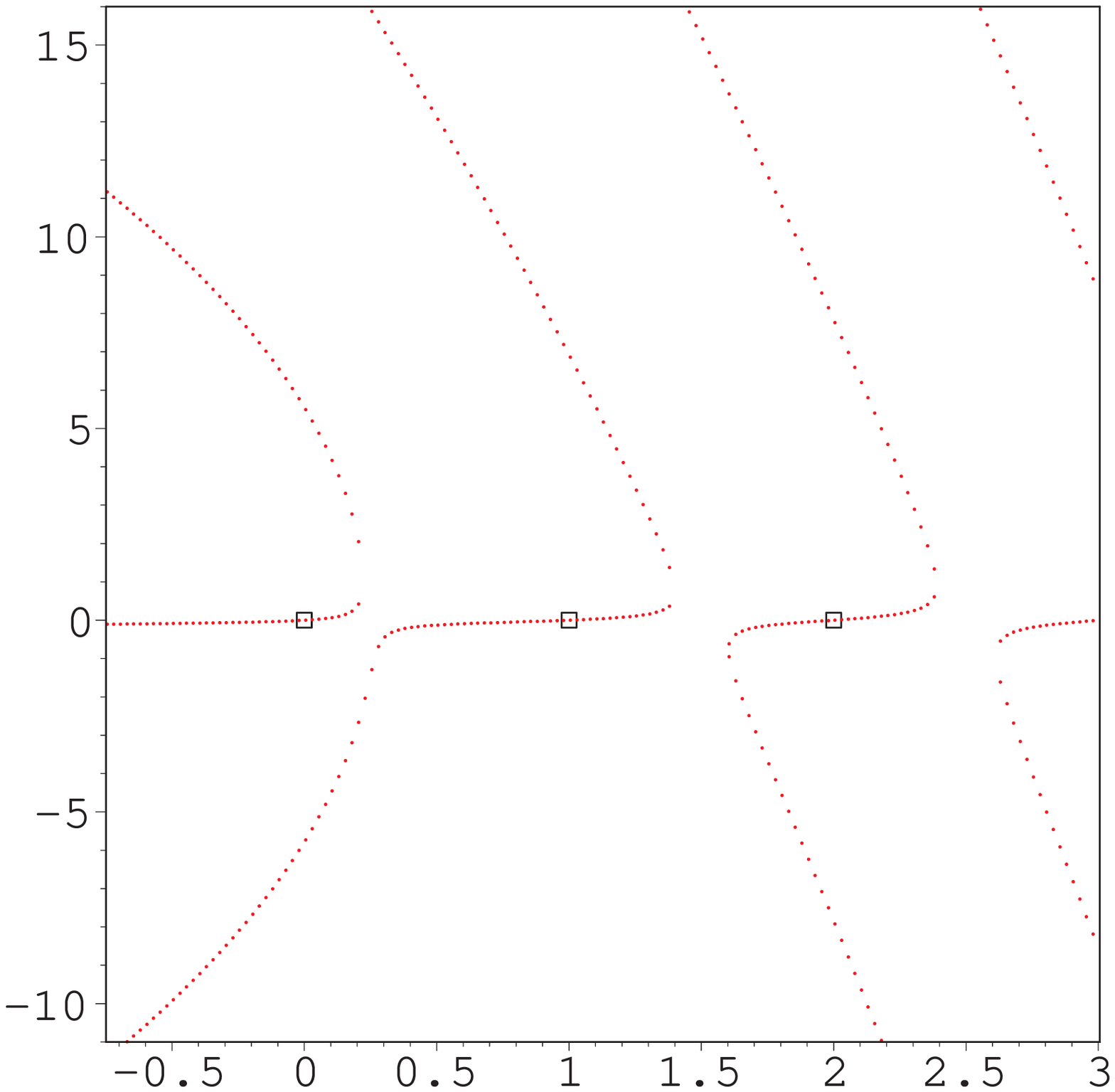}
\\[3pt]
\!\!\!\!\!\!
\parbox{0.32\linewidth}{\small~~~~~\,$\alpha_+=0.99$}&
\parbox{0.32\linewidth}{\small~~~~~\,$\alpha_+=1$}&
\parbox{0.32\linewidth}{\small~~~~~\,$\alpha_+=1.01$}
\end{array}
\]
\[
\parbox{0.77\linewidth}{\small\raggedright
Figure~\ref{ntrans}: Yet more complex eigenvalues for $M=3$, 
this time near the line
$\alpha_+=1$\,. Labelling as
in figure~\ref{trans}.}
\]

\noindent
Three of the low-lying levels in the middle plot of figure \ref{ntrans}, two
of which become complex for $\alpha_- >1/4$, can in fact be found exactly: they are
the roots of 
\eq
P_3(E) = E^3 -32[ (5-3 J) \alpha_- + J - 2]E=0
\label{p3}
\en
at $J=3$. 
More generally for $M{=}3$, at
$\alpha_+ =(J{-}1)/2$, 
$J=1,2,\dots$\,, $J$ levels of the 
${\cal PT}$-symmetric problem (\ref{PT}) can be found exactly. This can be
seen using the above-mentioned fifth spectral equivalence of \cite{DDT3}: on
the lines $\alpha_+=(J{-}1)/2$, 
(\ref{PT}) is mapped to the radial problem at a point where it is
quasi-exactly solvable \cite{Tur}\footnote{We remark that a relationship between
generalised supersymmetry and quasi-exact solvability, albeit for
a different
set of models, was originally discussed in \cite{KP}.}. 
(Equivalently, it follows directly from the termination of the relevant
Bender-Dunne \cite{BD} expansion.)
For $J$ odd, one of these QES levels is
the exactly-zero level mentioned in the previous paragraph.
This level (together with the other $J{-}1$ QES levels) can be eliminated by
passing from $(\alpha_+,\alpha_-)=((J{-}1)/2,\alpha_-)$ to 
$(\widehat\alpha_{+}, \widehat\alpha_{-})=(-(J{+}1)/2 ,\alpha_-{-}J/2)$ 
\cite{DDT3}, and the
exact locations of the level-crossings can again be found: they are at
$\alpha_-= J/2+n$, $n=0,1,\dots\,$. 
However, for the moment we do not know if a similar exact treatment can be
given for other values of $M$, though we expect the qualitative features of
the breaking of level-crossings to persist, at least for nearby values of $M$.
(Recall that the `protection' of the zero-energy state for
$\alpha_{\pm}\in\ZZ^+$ holds for all $M$.)

To end this section, we show three further plots which should help the reader to
understand how the levels reorganise passing between figures \ref{trans} and
\ref{ntrans}. The middle plot is another QES example.
This middle plot illustrates an additional feature of the models on the QES lines
$M{=}3$,
$\alpha_{\pm}=(J{-}1)/2$\,: the levels which become complex along these lines always
lie in the QES part of the spectrum. To prove this, we simply note that the
`dual' problems, with $(\alpha_{\pm},\alpha_{\mp})=
(\widehat\alpha_{\pm},\widehat\alpha_{\mp})
=(-(J{+}1)/2 ,\alpha_-{-}J/2)$, always
lie in regions of the parameter space covered by the reality proof of \cite{DDT3}.
As already mentioned, the spectrum of this dual problem is identical to that of
the original problem, minus the QES levels. (Strictly speaking, the arguments of \cite{DDT3}
applied to a radial problem, but the discussion of section
8 of that paper, based on intertwining operators, can be modified
to cover the current, lateral, problem. One can alternatively argue via the `fifth spectral
equivalence' mentioned at the start of this section.)
This demonstrates that the non-QES levels always remain real. A similar result was conjectured
for quartic QES potentials in \cite{BBQ}, but so far as we know remains unproven.
\[
\begin{array}{ccc}
\refstepcounter{figure}
\label{nntrans}
\!\!\!\!\!\!
\includegraphics[width=0.32\linewidth]{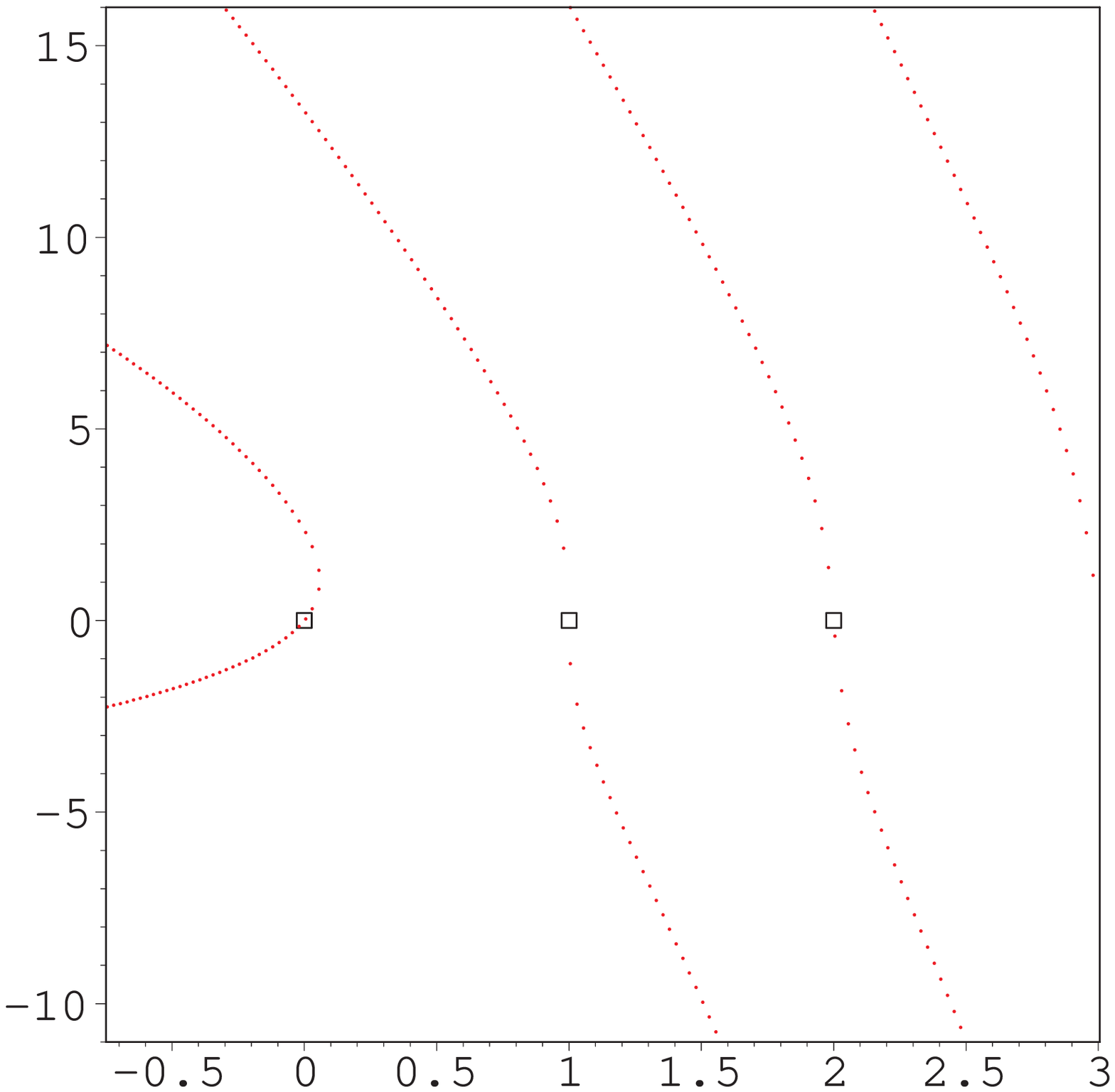}\,&
\includegraphics[width=0.32\linewidth]{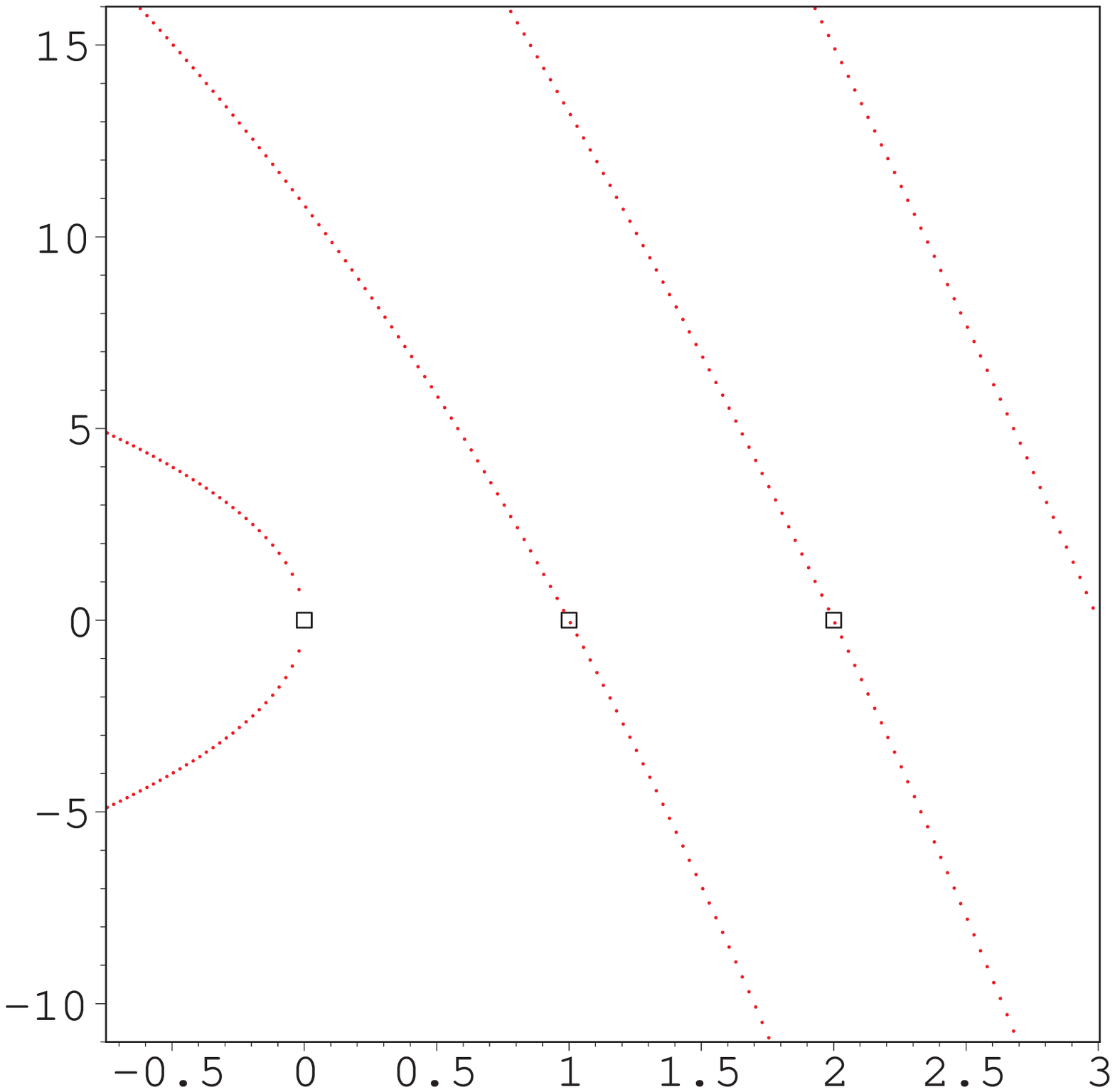}\,&
\includegraphics[width=0.32\linewidth]{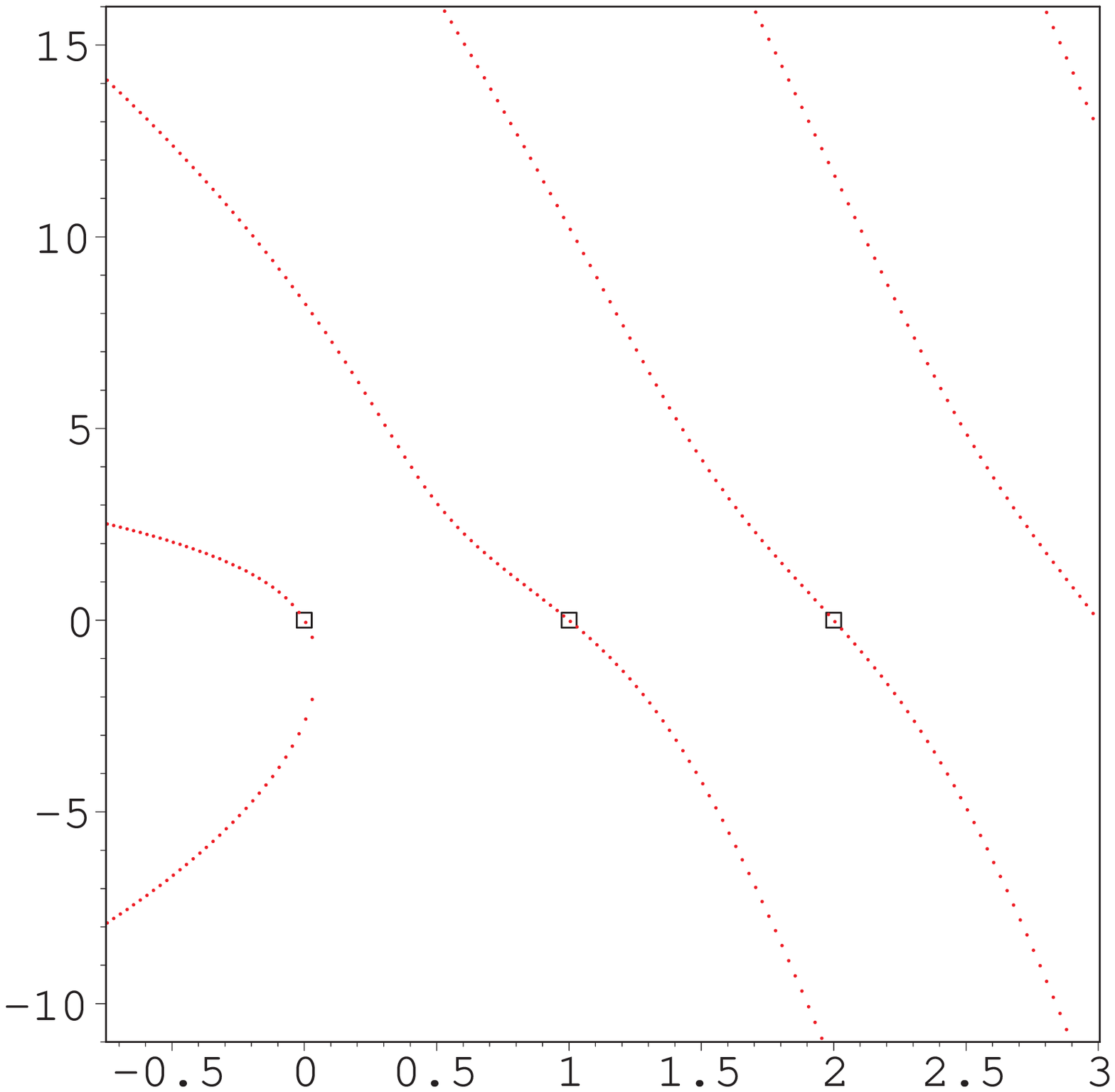}
\\[3pt]
\!\!\!\!\!\!
\parbox{0.32\linewidth}{\small~~~~~\,$\alpha_+=0.25$}&
\parbox{0.32\linewidth}{\small~~~~~\,$\alpha_+=0.5$}&
\parbox{0.32\linewidth}{\small~~~~~\,$\alpha_+=0.75$}
\end{array}
\]
\[
\parbox{0.77\linewidth}{\small\raggedright
Figure~\ref{nntrans}: The reorganisation of the levels passing from 
$\alpha_+=0$ to
$\alpha_+=1$\,. 
Labelling as
in figure~\ref{trans}.}
\]

\medskip

\medskip\noindent{\bf 4.}
The above discussion has confirmed 
that the domains where the spectrum of
(\ref{PT}) has a complex component
open out from the level-crossing points on the boundary
of the region $A$. These points are given exactly, for general $M$, by equation
(\ref{levcr}). It
is interesting to see, at a qualitative level,
how these `domains of unreality' coalesce as one moves further into $A$.
Figure \ref{scan} below exhibits some initial numerical results, found again
for $M=3$ using the {\sc Maple} code of \cite{DDT3}. This should be viewed as
a refinement of the initial `phase diagram' of figure~\ref{regions}.
The full domain of unreality is the interior of the curved line, a proper
subset of $A$ which only touches its boundary at the points
$(\alpha_{\pm},\alpha_{\mp})=(0,n{+}1/2)$, which are the level-crossing points
(\ref{levcr}) for $M{=}3$. 
In the small, approximately-triangular region inside $A$ but 
outside the
curved line which abuts the point $\alpha_+=\alpha_-=0$, the spectrum
is not only real but also entirely positive, despite the fact that it lies
outside the domain $D$. This shows that, while the condition 
$\alpha<M+1-|2l{+}1|$ is sufficient for positivity of the spectrum,
it is not necessary.

Perhaps the most striking feature of figure \ref{scan} is the pattern of cusps. 
In the absence of an analytical analysis, we do not yet know whether these are
special to $M=3$, or more generic. 
At least numerically, they lie exactly on the lines $\alpha_{\pm}=n$, 
along which the model possesses a protected zero-energy state (to
guide the eye, segments of these 
lines have also been added on figure \ref{scan}\,).
Furthermore, for $M=3$ the model is quasi-exactly
solvable on these lines. As shown above, the levels which go complex 
then lie in the QES part of the spectrum, and this allows the cusps 
to be located exactly for M=3, assuming that they do indeed lie on 
the QES lines. For example, this places the two lowest pairs of cusps at 
$(\alpha,l)=(9,-1/2{\pm}3/2)$ and $(15{-}3/\sqrt{2},-1/2{\pm}3/\sqrt{2})$\,.

\medskip

\[
\begin{array}{c}
\refstepcounter{figure}
\label{scan}
\!\!\!\includegraphics[width=0.42\linewidth]{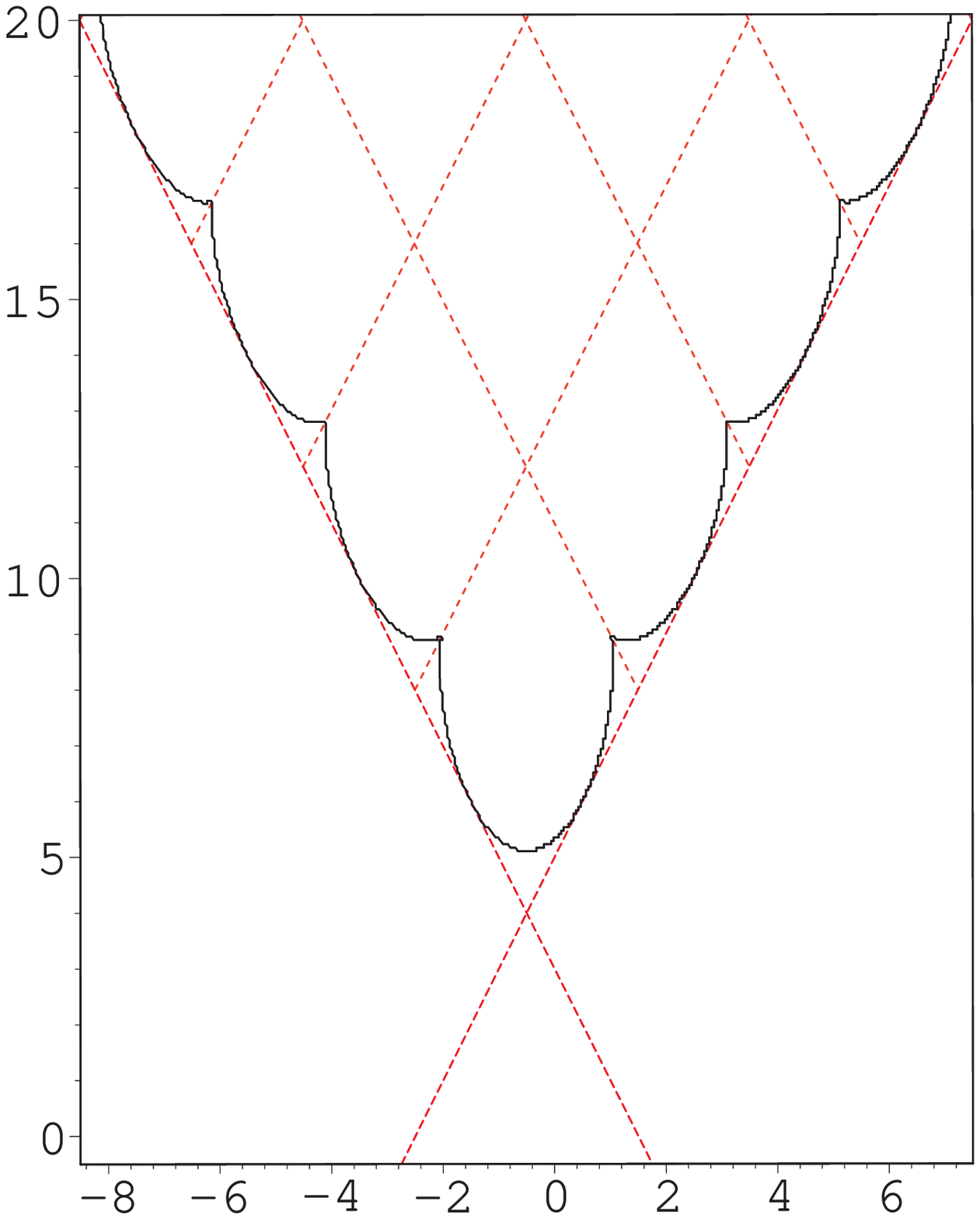}
\\[11pt]
\parbox{0.6\linewidth}{
\raggedright
{\small Figure \ref{scan}: The domain of unreality for $M=3$. Horizontal and
vertical axes are $l$ and $\alpha$ respectively, as in figure~\ref{regions}.
}}
\end{array}
\]

A detailed study of these properties must await future work;
we expect that complex WKB techniques 
will  be useful in this regard.
These were recently employed in \cite{BBMSS} for a problem
which can be mapped onto the $l=0$, $M=2$  case of
(\ref{PT}), save for a different choice of Stokes sectors --
$\CS_{-1}$ and $\CS_2$ -- for the quantisation contour.

The intersections of the lines $\alpha_+=(J_1{-}1)/2$
and $\alpha_-=(J_2{-}1)/2$
provoke one further thought, for $M{=}3$. At these points the model is quasi-exactly
solvable from two different points of view, with either $J_1$ or $J_2$ levels
lying in the QES part of the spectrum. This is reflected in a curious
factorisation property of the Bender-Dunne \cite{BD} polynomials
$P_J(\alpha_{\pm},E)$. (These
polynomials encode the QES levels in their
zeroes,  equation
(\ref{p3}) being one example.) Taking $J_1<J_2$, 
at the point
$(\alpha_+,\alpha_-)=((J_1{-}1)/2,(J_2{-}1)/2)$
where the two QES lines intersect,
$P_{J_1}(\alpha_-,E)$ is a factor of
$P_{J_2}(\alpha_+,E)$\,. A direct proof of this result, using
the Bender-Dunne three-term
recursion relation \cite{BD}, can also be given.

\medskip\noindent{\bf 5.}
We conclude with some general comments.
Our main purpose in this note has been to understand the physical reasons for the
breakdown in the reality proof of \cite{DDT3}. We have shown that, on the 
lines along which the proof fails,
the model has a hidden supersymmetry, and that once these lines are crossed,
the reality property can, and at some points does, fail to hold.
This implies that the conditions required by the proof are of more than
technical significance. It would be interesting to extend the proof to cover the
full domain for which the spectrum of (\ref{PTg}) is real, but the
complicated shape of the domain of unreality
shown in figure~\ref{scan} suggests that this will
be a difficult task.

The model we have been discussing has turned out to have an
unexpectedly rich structure, and should serve as a testing-ground for other
aspects of ${\cal PT}$-symmetric quantum mechanics. More detailed studies
should help us to gain a better understanding of the emergence of complex
eigenvalues in general, as well as shedding light on the pattern
of transitions revealed by figure \ref{scan}.  We should also remember that,
via the ODE/IM correspondence \cite{DTa,BLZa,DTb,Sc,DDTr},
all of these results have potential
implications in the field of integrable models in 1+1 dimensions.

Finally, we remark that similar appearances of complex levels preceded by
level-crossings
have been observed in a nonunitary model of quantum field theory -- the
boundary scaling Lee-Yang model \cite{DPTW}. 
In this case the level-crossings were protected by an identity between cylinder
partition functions in models with differing boundary conditions, which
followed from a set of functional equations called a $T$-system.
It remains to be seen whether this phenomenon can also be understood on the
basis of some hidden symmetry, as was the case for the level-crossings
discussed in a simpler, quantum-mechanical context in this letter.

\bigskip\noindent{\bf Acknowledgements --~}
We would like to thank G\'erard Watts for a useful discussion. 
PED thanks the EPSRC for an Advanced Fellowship, and
RT thanks the EPSRC for a VF grant, number GR/N27330.
%
%

\break

\end{document}